\documentclass[10pt]{iopart}

\newcommand{\mx}{\textrm{\scriptsize max}}
\newcommand{\bb}{\textrm{\scriptsize B}}
\newcommand{\LL}{\textrm{\scriptsize L}}

\newcommand{\vc}{\mathbf}
\usepackage{graphicx}
\usepackage{amssymb}
\usepackage{color}

\begin{document}

\title[Transverse Force]{Transverse Force Induced by a Magnetized Wake}

\author{Trevor Lafleur}
\address{PlasmaPotential-Physics Consulting and Research, Canberra, ACT 2601, Australia}

\author{Scott D.~Baalrud}
\address{Department of Physics and Astronomy, University of Iowa, Iowa City, Iowa 52242, USA}
\ead{scott-baalrud@uiowa.edu}

\begin{abstract}
The force on a test charge moving through a strongly magnetized plasma is calculated using linear response theory. 
Strong magnetization is found to generate a component of the force perpendicular to the velocity of the particle in the plane formed by the velocity and magnetic field vectors. 
This transverse force is generated by an asymmetry with respect to the velocity vector in the induced electrostatic wake potential that is associated with the action of the Lorentz force on the background plasma. 
The direction depends on the speed of the test charge. 
If it is faster than a critical speed characteristic of the sound speed, it acts to reduce the component of velocity parallel to the magnetic field and to increase the gyroradius. 
In contrast, if the speed is below this critical speed, it acts to increase the velocity parallel to the magnetic field and to decrease the gyroradius. 
Because the transverse force is perpendicular to the velocity, it does not directly influence the total energy of the test charge. 
Nevertheless, it significantly alters the trajectory on a timescale associated with the Coulomb collision time. 

\end{abstract}

%
%
\submitto{\PPCF}
%
%
\ioptwocol

\section{Introduction}

The trajectory of an individual charged particle in a plasma is a fundamental process.
It is of direct interest in many applications, such as fusion energy, because it describes the energy deposition rate of fusion products~\cite{sigm:71,fren:19} and the potential for runaway electrons~\cite{paz:14}. 
It also underlies kinetic theories used to describe macroscopic transport properties of plasmas. 
The force balance describing this trajectory usually consists of the Lorentz force acting on the charge due to an external magnetic field, $q_t \vc{v} \times \vc{B}$, and a friction force that acts in the direction opposing the particle velocity $-\nu \vc{v}$. 
In a fully ionized plasma, the friction is due to Coulomb collisions. 
It can be accurately described by linear response theory as the force arising from the induced electric field in the wake of the charged particle~\cite{nich:83,ichi:92}. 
For example, this concept underlies the Lenard-Balescu kinetic theory~\cite{lena:60,bale:60}. 
Here, it is shown that if the background plasma is strongly magnetized, an additional friction force arises that is in the direction perpendicular to $\vc{v}$ in the plane of $\vc{v}$ and $\vc{B}$.  

This transverse component of the force becomes appreciable when the gyrofrequency of the background species responsible for the drag exceeds its plasma frequency: $\omega_{c}/\omega_{p} \gg 1$, where $\omega_c = |q|B/m$ and $\omega_p = \sqrt{q^2n/\epsilon_om}$. 
In this situation, the equation of motion for the test charge is 
\begin{equation}
\label{eq:eom}
m_t\frac{d\vc{v}}{dt} = q_t \vc{v} \times \vc{B} - m_t \nu \vc{v} - m_t \nu_\times\, \vc{v} \times \hat{\vc{n}}  
\end{equation} 
where $\hat{\vc{n}}$ is the unit vector perpendicular to the plane containing $\vc{v}$ and $\vc{B}$: $\hat{\vc{v}} \times \hat{\vc{b}} = \sin \theta \hat{\vc{n}}$. 
Here, $\theta$ is the angle between $\vc{v}$ and $\vc{B}$, while $\hat{\vc{v}}=\vc{v}/v$ and $\hat{\vc{b}} = \vc{B}/B$ denote unit vectors. 
The term $-m_t\nu\vc{v}$ represents the usual component of the friction force, which acts to slow the particle, whereas the term $-m_t\nu_\times \vc{v} \times \hat{\vc{n}}$ represents the transverse component of the force, which is perpendicular to $\vc{v}$ but in the plane of $\vc{v}$ and $\vc{B}$. 
The transverse force vanishes if the particle velocity is either parallel ($\theta=0^\circ$) or perpendicular ($\theta = 90^\circ$) to the magnetic field, but can be large otherwise. 
It arises from the way in which the Lorentz force influences the induced charge distribution excited by the test charge. 
It will be shown that strong magnetization visibly distorts the wake potential of the test charge, generating an asymmetry in the induced charge density distribution that leads to the transverse force. 

We concentrate on conditions at which random thermal motion of the test charge, i.e., Brownian motion, is negligible. 
This is sometimes referred to as the field-particle component of the Coulomb collision rate. 
We also focus on the limit in which the background plasma can be treated as a one-component plasma (OCP), although it should be highlighted that this limit is not necessary for the formation of the transverse force.
This limit is relevant to many applications of interest, such as slowing of fusion products on electrons, slowing of fast (runaway) electrons on the thermal electrons, and electron cooling of heavy ion beams~\cite{ners:07}. 
In these cases, ions do not contribute to the linear wake because they are much more massive than electrons.  

Results of linear response theory predict that the sign of the transverse force depends on the speed of the test charge. 
If it is faster than a critical speed that is approximately the thermal speed of the background plasma, $v > v_c \sim v_T$, the force acts to decrease the speed of the test charge parallel to the magnetic field and to correspondingly increase its perpendicular speed. 
This increases its gyroradius. 
In contrast, if $v< v_c$, the force acts to increase its parallel speed and decrease its perpendicular speed. 
This decreases its gyroradius. 
These effects can significantly influence the overall trajectory and even the range of the test charge.
Since the transverse force is perpendicular to $\vc{v}$ it does not directly influence the energy loss rate. 

The influence of a strong magnetic field on the induced force has previously been studied using linear response theory~\cite{may:70,ware:93,joos:15,ners:00,cere:00,cere:05}, binary collision theory~\cite{ners:03,ners:09,ners:19}, a Green-Kubo formalism~\cite{deut:08}, classical trajectory Monte Carlo simulation~\cite{ners:07}, and particle-in-cell simulations~\cite{boin:96,hu:09}; see~\cite{ners:07} for a recent review. 
These works have shown that strong magnetization can significantly influence stopping power, which is associated with the component of the force along the velocity vector: $\vc{F}_v = F_v \hat{v} = - m_t \nu \vc{v}$. 
A series of work by Nersisyan \emph{et al}~\cite{ners:07,ners:00,ners:03,ners:09,ners:19} has provided a detailed analysis, including numerical solutions, and analytic approximations in a wide variety of limits associated with the particle speed, strength of the magnetic field, and angle between the velocity an magnetic field vectors. 

Recent work motivated by dust in magnetized plasmas~\cite{thom:12,boni:13} has detailed the influence of the magnetic field on the induced wake using linear response theory~\cite{joos:15}, particle-in-cell simulations~\cite{milo:17,milo:18,sund:18,dari:19}, molecular dynamics simulations~\cite{piel:18,piel:18b}, and experiments~\cite{jung:18}.
For example, Joost \emph{et al}~\cite{joos:15} have provided detailed computational solutions for the wake potential also using linear response theory, as well as including the influence of neutral collisions on the wake. 
Recent research has also used particle-in-cell simulations to explore the wake potential when the magnetic field is perpendicular to the test charge velocity~\cite{sund:19}. 
Other related topics in which similar calculations have been made are the wake induced by probes in magnetized plasmas~\cite{rubi:82,pata:07}, and the wake surrounding objects in space plasmas, such as the moon in the solar wind~\cite{hutc:12}. 
Our work may contribute to these studies by demonstrating the existence of a transverse component of the induced force when the flow is oblique to the magnetic field.

One shortcoming of using linear response theory to compute the force on a test charge is that it does not account for close collisions. 
As a result, the theory nominally diverges logarithmically in the limit of short distances (large wavenumber). 
The standard approach to resolve this divergence is to model close collisions by limiting the integrand to the thermal-averaged distance of closest approach in a binary collision, i.e., the Landau length $r_\LL = |q_tq|/(4\pi\epsilon_ok_\bb T)$~\cite{nich:83,ichi:92}. 
This same cutoff is expected to apply if the plasma is magnetized as long as the Landau length is much smaller than the gyroradius $r_c = v_\perp/\omega_c$. 
Previous work has explored binary collisions at conditions for which the gyroradius is smaller than the distance of closest approach~\cite{onei:83,glin:92}, but in this paper we limit the discussion to plasmas in which $r_c \gg r_\LL$. 
Since this condition is easily satisfied in most weakly coupled plasmas, this does not significantly limit the applicability of the theory. 
The exceptions occur in the transition to strong Coulomb coupling~\cite{baal:17}, as well as non-neutral plasma experiments that access the regime in which $r_c \ll r_\LL$~\cite{dubi:98,dubi:14}. 
In terms of length scales, magnetization is found to influence the wake when the gyroradius is smaller than the Debye length (corresponding to the timescale ordering of $\omega_c > \omega_p$). 
Thus, the target regime of interest in this work is that classified as ``strongly magnetized'' in~\cite{baal:17}: $r_\LL \ll r_c \ll \lambda_D$. 

We also note that nonlinear effects are known to influence wake potentials~\cite{hutc:11}. 
These may influence the force induced by the charge density of the wake. 
Further research will be needed to quantify the magnitude of nonlinear effects. 
Nevertheless, the basic mechanism of a transverse force associated with the wake potential is expected to apply even when nonlinear effects become important. 

The remainder of this paper is organized as follows. 
Section~\ref{sec:lrt} presents the linear response theory describing the wake induced by a test charge, and the associated force, in a strongly magnetized plasma. 
Section~\ref{sec:comp} explains the computational method that was used to solve for the wake potential and force. 
Section~\ref{sec:force} presents solutions for the induced force for a variety of speeds, orientations and magnetic field strengths. 
Section~\ref{sec:wake} presents results and a discussion of the wake potential surrounding the test charge. Section~\ref{sec:trajectories} describes example trajectories for a test case of a He$^{2+}$ ion slowing on an electron OCP.

\section{Linear Response Theory\label{sec:lrt}} 

A test charge moving through a plasma excites a charge density distribution in its wake. 
The force acting back on the test charge due to the induced electric field can be computed using the linearized Vlasov equation and Gauss's law. 
The associated wake potential is~\cite{nich:83,ichi:92,dewa:12}  
\begin{equation}
\label{eq:phi}
\phi(\vc{r},t) = \frac{q_t}{\epsilon_o} \int \frac{d^3k}{(2\pi)^3} \frac{e^{i \vc{k} \cdot (\vc{r} - \vc{v}t)}}{k^2 \hat{\varepsilon}(\vc{k}, \vc{k} \cdot \vc{v})} 
\end{equation}
where $\hat{\varepsilon}$ is the linear dielectric response function and $q_t$ is the magnitude of the test charge moving with velocity $\vc{v}$. 
In a magnetized plasma, this can be expressed as 
\begin{equation}
\label{eq:ep_integral}
\hat{\varepsilon}(\vc{k}, \omega) = 1 + \frac{q^2}{\epsilon_o k^2 m} \int d^3v^\prime \frac{\vc{k} \cdot \partial f/\partial \vc{v}^\prime}{\bar{\omega}} 
\end{equation}
where 
\begin{equation}
\label{eq:omega_s}
\frac{1}{\bar{\omega}} = -i \int_0^\infty d\tau\, \exp [ i(\vc{k} \cdot \vc{d} + \omega \tau)]
\end{equation}
is a defined frequency associated with the gyro-position   
\begin{eqnarray} 
\vc{d} &=& r_c^\prime [\sin \phi^\prime - \sin (\phi^\prime + \omega_{c} \tau)] \hat{x} \\ \nonumber
& & - r_c^\prime [\cos \phi^\prime - \cos(\phi^\prime + \omega_{c} \tau)] \hat{y} -  v_z^\prime \tau \hat{z} .
\end{eqnarray}
Here, $\vc{B} = B \hat{z}$ and $r_c^\prime = v_\perp^\prime/\omega_{c}$ is the gyroradius associated with the single-component background plasma. 
We concentrate on the case where $f$ is Maxwellian, so the dielectric function can be expressed as~\cite{kral:69} 
\begin{eqnarray}
\label{eq:full_epsilon}
& &\hat{\varepsilon}(\vc{k}, \omega) = 1 + \frac{1}{k^2 \lambda_{D}^2} \biggl[ 1 + \\ \nonumber
& & \frac{\omega}{|k_\parallel| v_{T}} \exp \biggl(- \frac{k_\perp^2 v_{T}^2}{2\omega_{c}^2} \biggr) \sum_{n=-\infty}^\infty I_n \biggl(\frac{k_\perp^2 v_{T}^2}{2\omega_{c}^2} \biggr) Z \biggl( \frac{\omega - n \omega_{c}}{|k_\parallel| v_{T}} \biggr) \biggr]
\end{eqnarray} 
where $I_n$ is the $n$-th order modified Bessel function of the first kind, $Z$ is the plasma dispersion function~\cite{frie:61}, and $k_{\parallel}$ and $k_{\perp}$ are wavenumbers parallel and perpendicular to the magnetic field direction respectively. 

The force on the test charge $\vc{F} = - q_t \nabla (\phi - \phi_C) |_{\vc{r}=\vc{v}t}$ can be obtained from the gradient of equation~(\ref{eq:phi}). 
Here, $\phi - \phi_C$ is the electrostatic potential of the induced charge distribution, and $\phi_C = q_t/(4\pi \epsilon_o |\vc{r}|)$ is the Coulomb potential associated with the test charge itself.
Since there is no self-force from the test charge in electrostatics, the force on the test charge is
\begin{equation}
\label{eq:force}
\vc{F} = - \frac{q_t^2}{\epsilon_o} \int \frac{d^3k}{(2\pi)^3} \frac{i \vc{k}}{k^2 \hat{\varepsilon}(\vc{k}, \vc{k} \cdot \vc{v})}  .
\end{equation} 
By applying the property that for a real argument ($\omega = \vc{k} \cdot \vc{v}$), the real component of $\hat{\varepsilon}(\vc{k}, \omega)$ is an even function of $\vc{k}$, the force can also be written~\cite{ichi:92} 
\begin{equation}
\label{eq:force_2}
\vc{F} = - \frac{q_t^2}{\epsilon_o} \int \frac{d^3k}{(2\pi)^3} \frac{\vc{k}}{k^2} \textrm{Im} \biggl\lbrace \frac{1}{\hat{\varepsilon}(\vc{k}, \vc{k}\cdot \vc{v})} \biggr\rbrace .
\end{equation} 
The remainder of this work will explore the solution of equation~(\ref{eq:force}) over a range of test particle velocities and magnetic field strengths. 

It is first noteworthy that the force vector can be expressed in two-dimensions because the component perpendicular to the $\vc{v}$--$\vc{B}$ plane vanishes. 
This can be seen by noting that equation~(\ref{eq:full_epsilon}) is an even function of $k_n$, where $k_n$ is the wavevector component in the direction perpendicular to the plane of $\vc{v}$ and $\vc{B}$ (the $\hat{\vc{n}}$ direction defined in the introduction). 
Thus, $\vc{F} \cdot \hat{\vc{n}} =0$ in equation~(\ref{eq:force}). 
Sometimes it is convenient to express the remaining two components as $\vc{F} = F_x \hat{x} + F_z \hat{z}$, where $\hat{z} = \vc{B}/B$ is the direction of the magnetic field and $\hat{x}$ is the direction perpendicular to $\vc{B}$ in the $\vc{v}$--$\vc{B}$ plane. 
In this expression, there will generally be two components of the force vector regardless of the strength of the magnetic field. 

Another convenient way to express the force is in terms of its components parallel and perpendicular to the velocity vector of the test charge: $\vc{F} = F_v \vc{v} + F_\times (\hat{\vc{v}} \times \hat{\vc{n}})$. 
The transformation between these coordinate systems can be made via the relations $\hat{\vc{v}} = \sin \theta \hat{x} + \cos \theta \hat{z}$, $\hat{\vc{n}} = - \hat{y}$, and $\hat{\vc{v}} \times \hat{\vc{n}} = \cos \theta \hat{x} - \sin \theta \hat{z}$. 
This representation highlights that a sufficiently strong magnetic field generates a force in the $\hat{\vc{v}} \times \hat{\vc{n}}$ direction that is not present in an unmagnetized plasma. 
In this representation, the component along $\vc{v}$ is
\begin{equation}
\label{eq:f_v}
F_v = -  \frac{q_t^2}{\epsilon_o} \int d\Omega^\prime \int_0^{k_\mx} \frac{dk}{(2\pi)^3} \frac{\vc{k} \cdot \hat{\vc{v}}}{k^2}  \textrm{Im} \biggl\lbrace \frac{1}{\hat{\varepsilon}(\vc{k}, \vc{k}\cdot\vc{v})} \biggr\rbrace,
\end{equation}
where $d\Omega^\prime \equiv \sin \theta^\prime d\theta^\prime d\phi^\prime$ is the solid angle expressing the angular component of $\vc{k}$ in a spherical coordinate system. 
This component of the force is commonly referred to as the stopping power. 
The transverse force is perpendicular to the stopping power, and can be expressed as 
\begin{equation}
\label{eq:f_times}
F_\times = - \frac{q_t^2}{\epsilon_o} \int d\Omega^\prime \int_0^{k_\mx} \frac{dk}{(2\pi)^3} \frac{k_\times}{k^2}  \textrm{Im} \biggl\lbrace \frac{1}{\hat{\varepsilon}(\vc{k}, \vc{k}\cdot\vc{v})} \biggr\rbrace.
\end{equation}
where $k_\times \equiv \vc{k} \cdot (\hat{\vc{v}} \times \hat{\vc{n}})$.
This expression for the transverse force is useful for illustrating the symmetries that it vanishes if the test charge is aligned either parallel ($\theta = 0^\circ$) or perpendicular ($\theta = 90^\circ$) to the magnetic field. 
For parallel alignment ($\theta = 0^\circ$) $\vc{k} \cdot (\hat{\vc{v}} \times \hat{\vc{n}}) = k_x = k \sin \theta^\prime \cos \phi^\prime$ and $\vc{k} \cdot \vc{v} = k_\parallel v_\parallel$, so the dielectric function (from equation~\ref{eq:full_epsilon}) is independent of $\phi^\prime$ and the overall $\phi^\prime$ integral vanishes. 
For perpendicular alignment ($\theta = 90^\circ$) $\vc{k} \cdot (\hat{\vc{v}} \times \hat{\vc{n}}) = -k_\parallel$ and $\vc{k} \cdot \vc{v} = k_\perp v_\perp$, so the dielectric function is even in $k_\parallel$ and the overall integrand of equation~(\ref{eq:f_times}) is odd in $k_\parallel$; so it too vanishes. 
Thus, the transverse force contributes only at oblique angles ($\theta \neq 0^\circ$, $90^\circ$, $180^\circ$, or $270^\circ$). 

One shortcoming of linear response theory in describing the influence of Coulomb collisions is that it does not account for strong (nonlinear) interactions that occur when particles get close to one another. 
In fact, the neglect of close collisions leads to a well-known logarithmic divergence when evaluating the force. 
The limit of close interactions are ``resolved'' by restricting the range of the $k$-integral (when using a spherical coordinate system) in equations~(\ref{eq:force})--(\ref{eq:f_times}) to be the inverse distance of closest approach in a binary collision~\cite{ichi:92} 
\begin{equation}
\label{eq:k_max}
k_{\mx} = 4\pi \epsilon_o \frac{\mu (v^2 + v_T^2)}{2|q q_t|} 
\end{equation}
where $\mu = m_tm/(m_t + m)$ is the reduced mass. 
Equation~(\ref{eq:k_max}) will be applied throughout this paper. 
However, it is important to point out that this expression is motivated by physical arguments that do not make reference to the magnetic field when considering the binary collision dynamics. 
This is normally a good approximation as long as the gyroradius is much larger than the distance of closest approach. 
It is uncertain how this cutoff should be changed if the gyroradius is smaller than the distance of closest approach~\cite{ners:03,ners:09,glin:92,hu:02}. 
Thus, it should be understood that the results are limited to the regime $r_c \gg r_\LL$. 

Throughout this paper we consider a background plasma that can be approximated as a one component plasma. 
In this limit, the equations can be expressed in terms of the Coulomb coupling parameter 
\begin{equation}
\Gamma = \frac{q^2/a}{4\pi\epsilon_o k_\bb T}
\end{equation}
the magnetization parameter 
\begin{equation}
\beta = \frac{\omega_c}{\omega_p}
\end{equation}
the Mach number $M \equiv v/v_T$, the angle between the velocity and magnetic field vectors $\theta$, the charge ratio $q_t/q$ and the mass ratio $m_t/m$. 
Here, $a = (3/4\pi n)^{1/3}$ is the average distance between particles and $\omega_c$ and $\omega_p$ are the angular gyrofrequency and plasma frequency associated with the OCP. 
By taking the Debye length to be the unit of distance, the dimensionless terms that arise in the dielectric response function can be written as
\begin{equation} 
\frac{k_\perp^2 v_T^2}{2\omega_c^2} = \frac{(k_\perp \lambda_D)^2}{\beta^2} \ \ \ \textrm{and} \ \ \
\frac{|k_\parallel | v_T}{\omega_c} = \frac{|k_\parallel \lambda_D|}{\sqrt{\beta/2}} .
\end{equation}
In these dimensionless units, the force is naturally expressed in term of $(q_t/q)^2\Gamma^2 k_\bb T/a$. 
When applying these units, the only place that the Coulomb coupling strength ($\Gamma$) enters is through the $k_\mx$ cutoff 
\begin{equation}
k_\mx \lambda_D = \frac{1}{\sqrt{3} \Gamma^{3/2}} \biggl| \frac{q}{q_t} \biggr| \frac{\mu}{m} \biggl(1 + \frac{v^2}{v_T^2} \biggr) .
\end{equation}
In terms of dimensionless variables, the requirement that $r_\LL \ll r_c$ implies that $\beta \ll \Gamma^{-3/2}$. 

Finally, we point out two limits of these expressions: zero magnetic field and infinite magnetic field. 
In the $B \rightarrow 0$ limit, the dielectric function can be expressed in the well-know form
\begin{equation}
\label{eq:beta_0}
\hat{\varepsilon}_{0} (\vc{k},\omega) = 1 - \frac{\omega_{p}^2}{k^2 v_{T}^2} Z^\prime \biggl( \frac{\omega}{kv_{T}} \biggr) .
\end{equation}
Since this depends on $\omega = \vc{k} \cdot \vc{v}$ and $k^2$, the transverse force from equation~(\ref{eq:f_times}) vanishes by the same arguments used to show that it vanishes for parallel or perpendicular orientation above. 
In the $B \rightarrow \infty$ limit the $n=0$ Bessel function term from equation~(\ref{eq:full_epsilon}) dominates and the result simplifies to
\begin{equation}
\label{eq:beta_inf}
\hat{\varepsilon}_{\infty} (\vc{k}, \omega) = 1 + \frac{1}{k^2 \lambda_{D}^2} \biggl[ 1 + \frac{\omega}{|k_\parallel | v_{T}} Z \biggl( \frac{\omega}{|k_\parallel|v_{T}} \biggr) \biggr] .
\end{equation} 
This limit is much less computationally expensive to evaluate than equation~(\ref{eq:full_epsilon}), and will be explored extensively below.

\section{Computational Method\label{sec:comp}}

\subsection{Force calculation} 

\subsubsection{Gordeev integral representation} 

Evaluation of the force components for an arbitrarily magnetized background plasma is non-trivial because of the complicated nature of the linear plasma dielectric response function, equation~(\ref{eq:full_epsilon}), as well as numerical challenges associated with evaluating the resulting force integrals in equation~(\ref{eq:force}). 
Rather than sum Bessel functions in equation~(\ref{eq:full_epsilon}), it often proved to be computationally expeditious to write the linear dielectric response function in terms of the Gordeev integral \cite{cavalier:13}  
\begin{eqnarray}
\label{eq:gordeev}
\hat{\varepsilon}(\vc{k}, \omega) &=& 1 + \frac{1}{k^2 \lambda_{D}^2} \biggl[1 + \\ \nonumber
& & iA\int_0^{\infty}dx e^{-B(1 - \cos x) - \frac{1}{2}Cx^2 + iAx} \biggr]
\end{eqnarray} 
where $A = \omega/\omega_c = {\bf k}\cdot{\bf v}/\omega_c$, $B = k_{\perp}^2v_T^2/2\omega_c^2$, and $C = k_{\parallel}^2v_T^2/2\omega_c^2$. Written in this form, the infinite summation and special functions are replaced by an integral with an integrand involving only elementary functions. 
The integral was computed using an adaptive form of Simpson's rule, which is rapid and allows for good numerical accuracy. 
The number of integration points was increased in cycles until the error in successive cycles fell below a chosen error tolerance. 
By choosing the integration step size correctly, function evaluations computed during previous cycles were reused in subsequent cycles.

The force components in equation~(\ref{eq:force}) were evaluated in the $x-z$ coordinate representation. 
Substituting equation~(\ref{eq:gordeev}) into (\ref{eq:force}), transforming to a spherical coordinate system such that, $d^3k = k^2\sin\theta^\prime dkd\theta^\prime d\phi^\prime$, $k_z = k\cos\theta^\prime$, $k_x = k\sin\theta^\prime \cos\phi^\prime$, $k_y = k\sin\theta^\prime \sin\phi^\prime $, $k_{\perp}^2 = k_x^2 + k_y^2 = k^2\sin^2\theta^\prime$, and simplifying, we obtain  
\begin{equation}
\label{eq:Fnorm_z}
\bar{F}_z = -\frac{6}{\pi^2} \int_0^{\pi/2}d\theta^\prime \sin\theta^\prime \cos\theta^\prime \int_0^{\pi}d\phi^\prime P(\theta^\prime, \phi^\prime)
\end{equation} 
and
\begin{equation}
\label{eq:Fnorm_x}
\bar{F}_x = -\frac{6}{\pi^2} \int_0^{\pi/2}d\theta^\prime \sin^2\theta^\prime  \int_0^{\pi} d\phi^\prime \cos\phi^\prime  P(\theta^\prime, \phi^\prime)
\end{equation} 
where
\begin{equation}
\label{eq:P}
P(\theta^\prime, \phi^\prime) \equiv \int_0^{\bar{k}_\mx} d\bar{k} \frac{\bar{k}^3 G_i}{G_r^2 + G_i^2} ,
\end{equation}
\begin{equation}
G_r = 1+ \bar{k}^2 - A\int_0^{\infty}dx\sin(Ax)e^{-B(1 - \cos x) - \frac{1}{2}Cx^2},
\end{equation} 
\begin{equation}
G_i = A\int_0^{\infty}dx\cos(Ax)e^{-B(1 - \cos x) - \frac{1}{2}Cx^2},
\end{equation} 
\begin{equation}
A = \frac{\sqrt{2} M \bar{k}}{\beta}\left(\sin\theta^\prime \cos\phi^\prime \sin\theta + \cos\theta^\prime \cos\theta \right),
\end{equation} 
\begin{equation}
B = \frac{\bar{k}^2\sin^2\theta^\prime}{\beta^2},
\end{equation}
\begin{equation}
C = \frac{\bar{k}^2\cos^2\theta^\prime}{\beta^2} ,
\end{equation} 
and the dimensionless quantities are $\bar{k} \equiv k \lambda_D$, and $\bar{F} \equiv F a q^2/(k_\bb T \Gamma^2 q_t^2)$ for each component of $F$. 
Equations~(\ref{eq:Fnorm_z}) and (\ref{eq:Fnorm_x}) have made use of the symmetry of the integrand, which allows the integration limits of each of the angular integrals to be halved. 
The stopping power and transverse force components are then found using a simple coordinate rotation
\begin{eqnarray}
F_v &=& F_x \sin \theta + F_z \cos \theta \\ 
F_\times &=& F_x \cos \theta - F_z \sin \theta .
\end{eqnarray}

The angular integrals in equations~(\ref{eq:Fnorm_z}) and (\ref{eq:Fnorm_x}) are straightforward and can be evaluated using typical numerical integration methods. 
The $\bar{k}$-integral however, is more complicated. As the Mach number of the test charge increases, the integrand becomes very sharply-peaked (approaching a Dirac delta function) so that the integration mesh may no longer correctly resolve it. Since this peak can be extremely narrow, simply increasing the number of integration points is not a feasible solution. 
To resolve this problem, we made use of a non-uniform integration mesh, together with a custom peak detection algorithm. 
The approximate position of this peak was first found by numerically determining the location of the local maxima of the integrand with respect to $k$. 
The integration mesh was then refined by adding a number, $N$, of new mesh points centered around the $k$-position of this maxima, $k_{\scriptsize \textrm{peak}}$, and spanning lower and upper bounds based on a chosen ``standard deviation'' ($k_{\scriptsize \textrm{width}}$, which was a parameter of the code). 
The above process was then successively repeated until the peak was completely resolved within some chosen tolerance. 
The use of lower and upper bounds provided a safety margin during each refinement, and ensured that the true peak location was not missed. 
Care was required when using the peak detection algorithm since for some parameters the integrand displays multiple peaks, only one of which was usually sharply-peaked. 
The numerical calculations were performed in Fortran, and computational speed was increased by making use of the OpenMP library.

\subsubsection{Strong field limit} 

In the strong field limit, $\beta \rightarrow \infty$, the dielectric can be modeled using equation~(\ref{eq:beta_inf}) and the $\bar{k}$ integral in equation~(\ref{eq:P}) can be evaluated analytically 
\begin{eqnarray}
\label{eq:P_inf}
P_\infty(\alpha, & \gamma) = \frac{\gamma}{4} \ln\left[\frac{\gamma^2 + \left(\alpha^2 + \bar{k}_{\mx}^2\right)^2}{\alpha^4 + \gamma^2}\right]  \\ \nonumber
&+ \frac{\alpha^2}{2} \left[\arctan\left(\frac{\alpha^2}{\gamma}\right) - \arctan\left(\frac{\alpha^2 + \bar{k}_{\mx}^2}{\gamma}\right)\right]
\end{eqnarray} 
where
\begin{equation}
\label{eq:alpha}
\alpha^2 \equiv 1 - 2\zeta D(\zeta) ,
\end{equation} 
\begin{equation}
\gamma^2 \equiv \pi\zeta^2 e^{-2\zeta^2} ,
\end{equation} 
$\zeta \equiv M\textnormal{sgn}(\cos\theta^\prime)\left(\cos\theta + \tan\theta^\prime \cos\phi^\prime \sin\theta\right)$ , and $D(\zeta)$ is the Dawson function, which is related to the plasma dispersion function as $Z(\zeta) = -2 D(\zeta) + i \sqrt{\pi} \exp(-\zeta^2)$. 
The components of the force in the strong magnetic field limit were computed by applying equation~(\ref{eq:P_inf}) in equations~(\ref{eq:Fnorm_z}) and (\ref{eq:Fnorm_x}) and computing the angular integrals numerically. 
This limit is much less computationally expensive to evaluate than for an arbitrary magnetic field strength because the analytic solution avoids the issues associated with a peaked $k$ integrand in equation~(\ref{eq:P}).

Note that if the test charge velocity vector is aligned with the magnetic field vector (i.e. $\theta = 0$), then $F_x = 0$, $F_{\times} = 0$, $F_v = F_z$, and the angular integrals in equation~(\ref{eq:Fnorm_z}) can be performed leading to the analytic solution 
\begin{equation}
\label{eq:Fnorm_z_inf_eta0}
\bar{F}_{z} |_{\beta \rightarrow \infty, \theta \rightarrow 0} = - \frac{3}{\pi} P_\infty (a,g)
\end{equation} 
where now $a^2 \equiv 1 - 2M D(M)$ 
and $g^2 = \pi M^2 e^{-2M^2}$.  
Equation (\ref{eq:Fnorm_z_inf_eta0}), written in a slightly different form, is identical to previously derived expressions in the literature (see for example Ref. \cite{ners:07}). 

\subsubsection{Unmagnetized limit} 

For an unmagnetised background plasma, the plasma dielectric function can be represented using equation~(\ref{eq:beta_0}). 
In this limit, the transverse force vanishes (as discussed above), and the stopping power can be expressed as
\begin{equation}
\label{eq:Fnorm_z_0}
\bar{F}_v |_{\beta=0} = -\frac{6}{\pi}\int_0^{\pi/2}d\theta^\prime \sin \theta^\prime \cos\theta^\prime P_o(\alpha_o, \gamma_o)
\end{equation} 
where $P_o(\alpha_o,\gamma_o)$ is defined the same way as $P_\infty(\alpha,\gamma)$ in equation~(\ref{eq:P_inf}), but with $\alpha \rightarrow \alpha_o$ and $\gamma \rightarrow \gamma_o$, where 
\begin{equation}
\alpha_o^2 \equiv 1 - 2M\cos\theta^\prime D(M\cos\theta^\prime)
\end{equation} 
and
\begin{equation}
\gamma_o^2 \equiv \pi M^2\cos^2\theta^\prime e^{-2M^2\cos^2\theta^\prime}.
\end{equation}

\subsection{Wake potential calculation} 

Computation of the electrostatic wake potential made use of two different algorithms depending on the nature of the background plasma. 
In both algorithms, the Fourier integrals in equation~(\ref{eq:phi}) were related to Discrete Fourier Transforms (DFTs), which were evaluated using a Fast Fourier Transform (FFT). 
For an unmagnetized plasma, the combination of equations~(\ref{eq:phi}) and~(\ref{eq:beta_0}) were solved using a similar numerical procedure to that described in references~\cite{joos:15} and \cite{ludwig:14}. 
In particular, a 2D FFT was performed in the plane perpendicular to $\vc{v}$, and 1D FFTs were performed along $\vc{v}$. 
This approach exploits the azimuthal symmetry of the problem, and only a single half-plane (parallel and perpendicular to $\vc{v}$) is needed to describe the entire 3D wake potential. 
The FFT algorithms used were taken from the SciPy library in Python, while population of the FFT arrays were performed in Fortran. 
For $M \gtrsim 0.5$, numerical artifacts associated with the peaked nature of the integrand in equation~(\ref{eq:phi}) were observed in the FFT. 
To remove these artifacts, a similar solution as \cite{joos:15,ludwig:14} was applied where a small amount of damping was added to the dielectric response function in the form of a BGK collision term. 
Comparison of results from the present code show excellent agreement with the wake potential results previously published in \cite{joos:15}.\footnote{In performing this comparison, a two-component plasma dielectric response function was used.}

Because of the high computational cost required to evaluate the plasma dielectric function for an arbitrary field strength, wake potentials were computed in the zero field limit of equation~(\ref{eq:beta_0}) and the strong field limit of equation~(\ref{eq:beta_inf}).
The comparison of these limits provides insight into how a strong magnetic field influences the wake potential. 

The strongly magnetized limit was evaluated by first substituting equation~(\ref{eq:beta_inf}) into equation~(\ref{eq:phi}) and taking the test charge to be at the origin of a Cartesian coordinate system (so $t=0$). 
The $k_y$ integral was evaluated analytically using the identity~\cite{mathematica} 
\begin{equation}
\int_{-\infty}^{\infty}dx \frac{e^{iy x}}{x^2 + z} = \frac{\pi}{\sqrt{z}}\left[e^{\sqrt{z} y}\Theta(-y) + e^{-\sqrt{z} y} \Theta(y)\right]
\end{equation}
where $\Theta$ is the Heaviside step function. 
With this, the dimensionless wake potential can be written 
\begin{equation}
\label{eq:phi_inf}
\bar{\phi} = \frac{\sqrt{3}}{2\pi^2} \int d^2\bar{k} e^{i(\bar{k}_x \bar{x} + \bar{k}_z \bar{z})} \frac{e^{c \bar{y}} \Theta (-\bar{y}) + e^{-c \bar{y}} \Theta (\bar{y})}{c}
\end{equation}
Here, $\bar{\phi}(\vc{r}) \equiv q^2 \phi(\vc{r})/(k_\bb T \Gamma^{3/2}q_t)$ is the dimensionless wake potential, $d^2\bar{k} \equiv d\bar{k}_x d\bar{k}_z$, 
\begin{equation}
c \equiv \sqrt{\bar{k}_x^2 + \bar{k}_z^2 + 1 + \zeta Z(\zeta)}
\end{equation}
and $\zeta \equiv M (k_z \cos \theta + k_x \sin \theta)/|k_z|$ is defined the same way as in equation~(\ref{eq:alpha}).

Equation~(\ref{eq:phi_inf}) was solved by again relating the continuous Fourier integrals to DFTs, and then making use of a 2D FFT. 
Because one of the integrals has been performed analytically (the $k_y$ integral), only a single 2D FFT is required for a given value of $y$. 
This allows for very rapid computations of the electrostatic wake potential, and significantly reduces memory management issues which would be required for a full 3D FFT. 
By evaluating equation~(\ref{eq:phi_inf}) at different values of $y$, the full 3D wake potential can be constructed.

\subsection{Trajectory calculation} 

To compute the test particle trajectories shown in section~\ref{sec:trajectories}, equation~(\ref{eq:eom}) (appropriately normalized) was solved using the force components for a very strongly magnetized plasma calculated from equations (\ref{eq:Fnorm_z}), (\ref{eq:Fnorm_x}) and~(\ref{eq:P_inf}). 
The computation used the LSODA routine~\cite{hind:83} that forms part of the odeint library in Fortran~\cite{ODE} (called using the SciPy library in Python~\cite{ODE_python}).

\section{Induced force\label{sec:force}} 

This section shows direct numerical solutions of the force on the test charge computed from equation~(\ref{eq:force}) for a range of speeds, angles and magnetic field strengths. 
These demonstrate that the magnetic field influences the force on the test charge when $\beta \gtrsim 1$ by both changing the magnitude of the stopping power component, $F_v$, as well as by introducing a transverse component $F_\times$ to the force vector.  
Furthermore, the force vector is well approximated by the $\beta \rightarrow \infty$ limit of the dielectric function from equation~(\ref{eq:beta_inf}) as $\beta$ approaches a value of approximately $\Gamma^{-3/2}$. 
Throughout this section, the coupling strength is taken to be $\Gamma = 10^{-3}$ when computing $k_\mx \lambda_D$ and the magnitude of the test charge is taken to be the same as the background particles $|q_t/q|=1$. 

\begin{figure}
\begin{center}
\includegraphics[width=8.0cm]{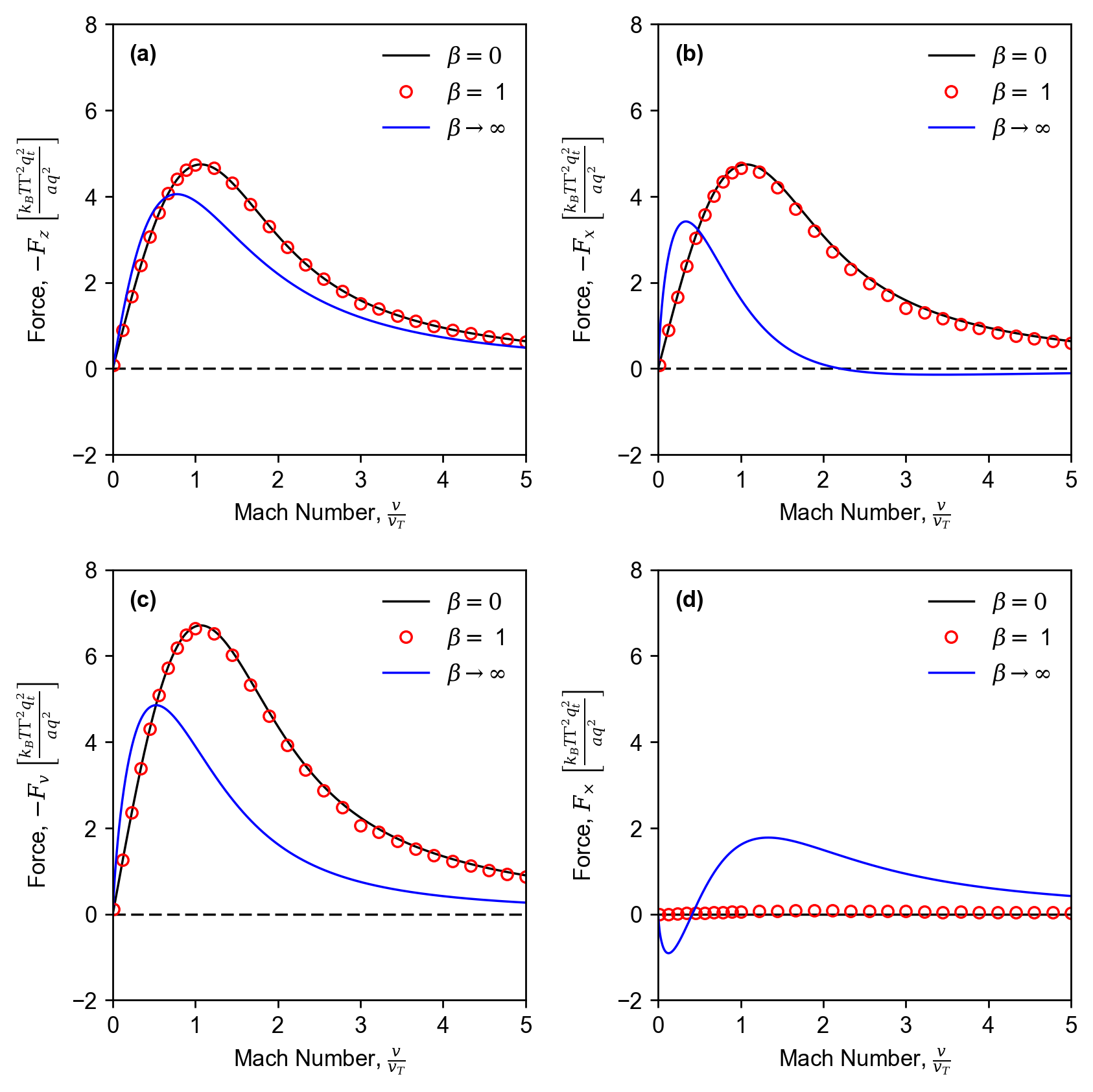}
\caption{Force on a test charge moving at $\theta = 45^\circ$ with respect to the magnetic field in a OCP with magnetization parameter $\beta = 1$ and coupling strength $\Gamma = 10^{-3}$ (circles). Two coordinate orientations are shown: (a)-(b) Parallel and perpendicular to the magnetic field, and (c)-(d) along the velocity vector and transverse.  Also shown are the unmagnetized (black lines) and infinitely magnetized (blue lines) limits. Note that (a)-(c) show the negative of the forces. }
\label{fg:force_beta1}
\end{center}
\end{figure}

Figure~\ref{fg:force_beta1} shows the force components in the case that $\beta = 1$ and $\theta = 45^\circ$. 
Components are shown both for the orientation with respect to the magnetic field ($F_x, F_z$) and with respect to the velocity vector ($F_v, F_\times$). 
The figure demonstrates that the magnetic field does not significantly influence the force at this value of $\beta$. 
Similar results are observed for any orientation of the velocity and magnetic field vectors, as well as for any value of $\beta$ less than unity. 
These observations suggests that the magnetic field does not significantly influence the friction force on a test charge if $\beta \lesssim 1$. 

\begin{figure}
\begin{center}
\includegraphics[width=8.0cm]{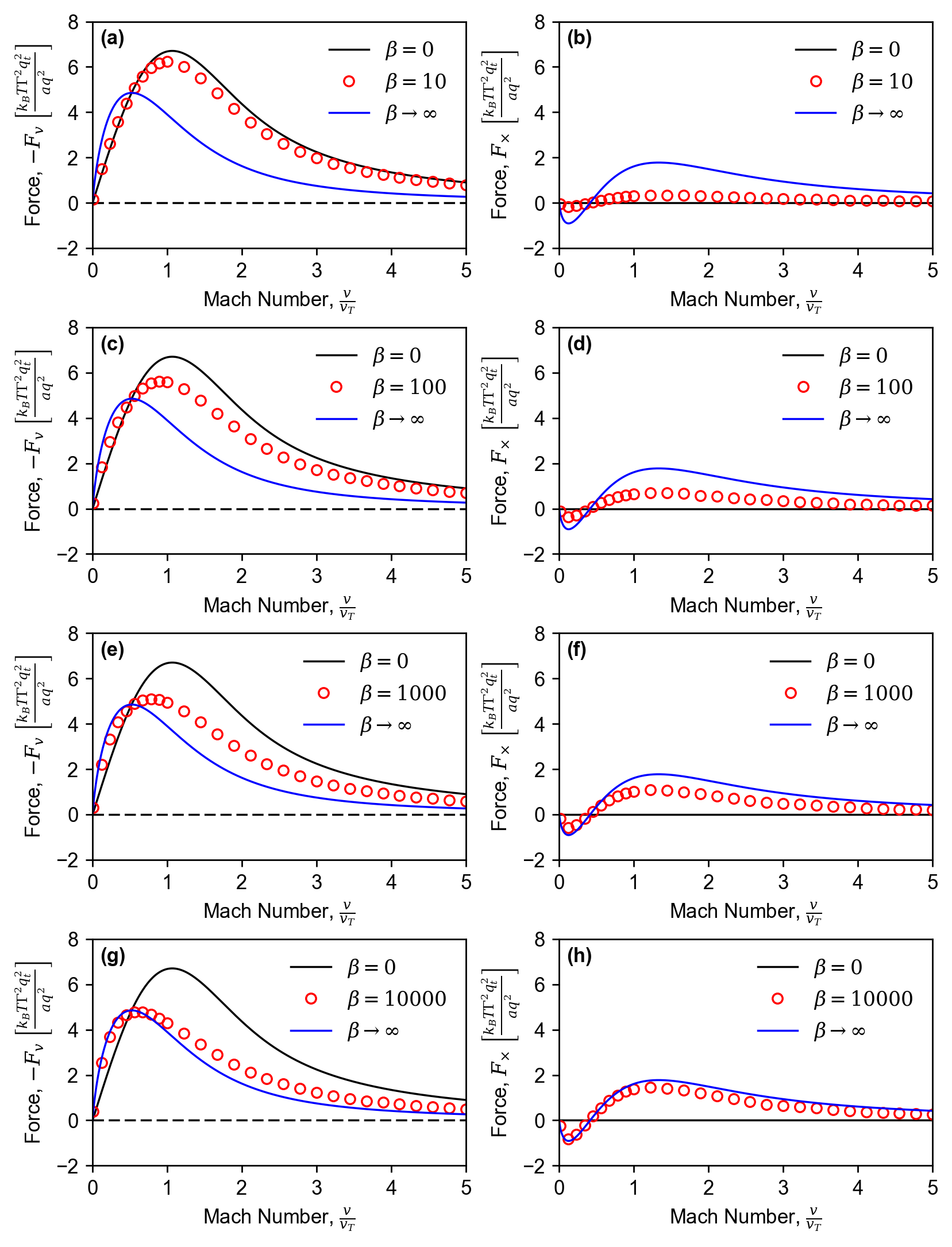}\\
\caption{Force on a test charge moving at a $45^\circ$ angle with respect to the magnetic field in a OCP with coupling strength $\Gamma = 10^{-3}$ and magnetization parameters (a),(b) $\beta = 10$, (c),(d) $\beta = 10^2$, (e),(f) $\beta=10^3$ and (g),(h) $\beta = 10^4$ (circles). Components are either parallel (left) or transverse (right) to the velocity vector.  Also shown are the unmagnetized (black lines) and infinitely magnetized (blue lines) limits. Note that the left panels show the negative of $F_v$.}
\label{fg:force_beta}
\end{center}
\end{figure}

Figure~\ref{fg:force_beta1} also shows that the predicted behavior in the strongly magnetized ($\beta \rightarrow \infty$) limit is very different from the unmagnetized limit. 
Since the angle is chosen to be $45^\circ$, $F_x=F_z$ in the unmagnetized limit. 
This corresponds to the usual expectation that the force is aligned along the velocity vector, so $F_v$ is non-zero but $F_\times = 0$, as indicated by the black line in the figure. 
Qualitatively different behavior is observed in the strongly magnetized limit. 
This limit is indicated by the blue curves, which correspond to the solution of equations~(\ref{eq:Fnorm_z}), (\ref{eq:Fnorm_x}) and~(\ref{eq:P_inf}). 
The magnitude of the force, as well as the position of the peak, is altered both parallel ($F_z$) and perpendicular ($F_x$) to the magnetic field, but in an asymmetric fashion with $F_x$ changing much more substantially than $F_z$. 
In the velocity-aligned coordinate system, the stopping power ($F_v$) is observed to decrease for high speeds, increase for low speeds, and the Bragg peak to shift to a lower speed in comparison with the unmagnetized limit. 
These effects have been observed in previous studies~\cite{ners:07}. 

\begin{figure}
\begin{center}
\includegraphics[width=7cm]{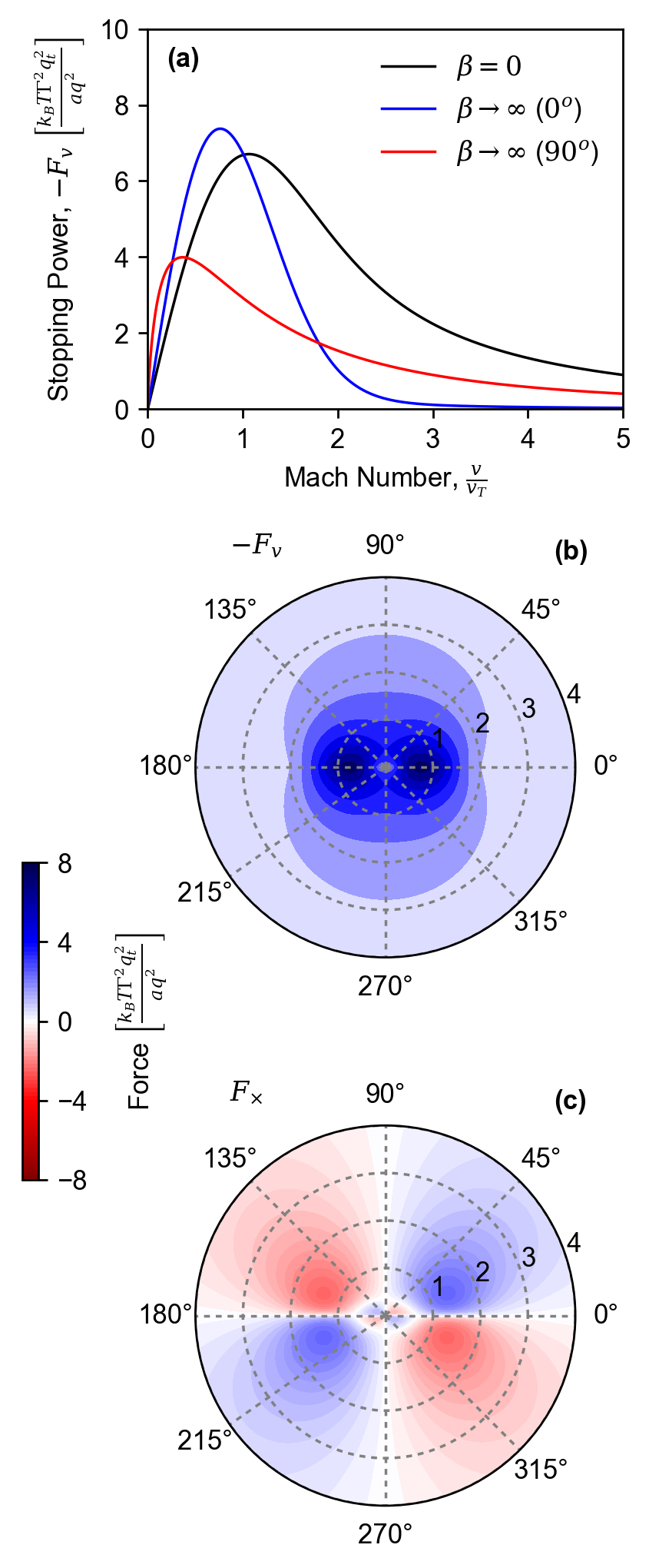}
\caption{(a) Stopping power ($-F_v$) of a test charge in an unmagnetized plasma (black line) compared with the $\beta \rightarrow \infty$ limit when the velocity of the charge is oriented parallel (blue line) or perpendicular (red line) to the magnetic field. The bottom panels show polar plots of the two force components computed in the $\beta \rightarrow \infty$ limit: (b) stopping power $(-F_v)$ and (c) transverse force $F_\times$. Here, the coupling strength is taken to be $\Gamma = 10^{-3}$.}
\label{fg:force_limits}
\end{center}
\end{figure}

The figure also makes a new observation that in addition to changing the stopping power, the strong magnetic field introduces a transverse component to the force. 
This transverse force is even larger than the stopping power in the high speed limit, and is generally non-negligible in magnitude at any speed. 
It is observed to switch sign at a speed of approximately $v_T/2$. 
A positive sign at a $45^\circ$ angle indicates that the transverse force acts to increase the gyroradius of the test charge, while a negative sign indicates that it decreases the gyroradius. 
We will see below that there is a phase dependence to the sign as well. 

\begin{figure}
\begin{center}
\includegraphics[width=8.0cm]{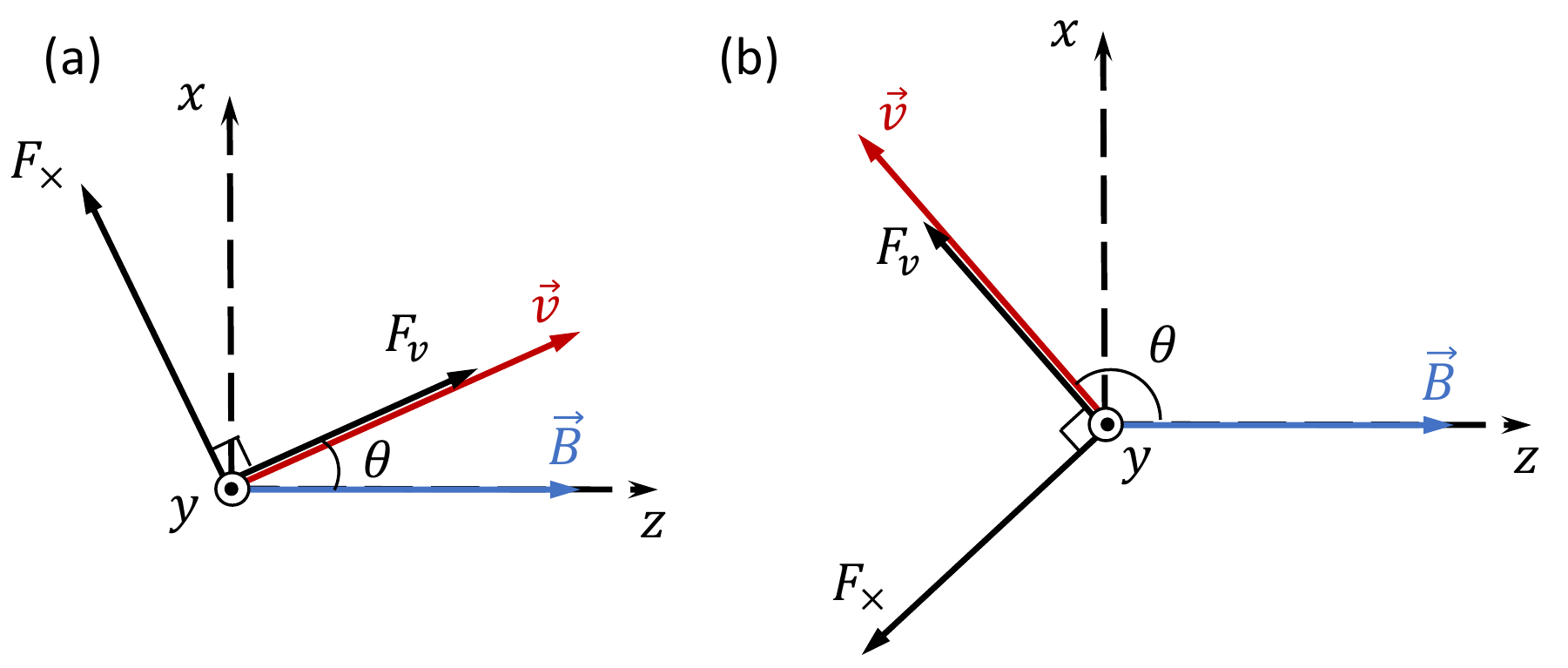}
\caption{Sketch of the force vectors in the $\vc{v}-\times$ coordinate system when the angle $\theta$ lies (a) in the first quadrant ($0 < \theta < 90^\circ$) and (b) in the second quadrant ($90^\circ < \theta < 180^\circ$).}
\label{fg:sketch}
\end{center}
\end{figure}

Figure~\ref{fg:force_beta} shows a series of solutions in the velocity-aligned coordinate system as $\beta$ increases from 10 to $10^4$. 
This shows that a significant influence of the magnetic field arises at $\beta = 10$, and that it smoothly transitions toward the $\beta \rightarrow \infty$ limit as the magnetic field strength increases.  
For $\beta = 10^4$, the full numerical solution closely approaches the $\beta \rightarrow \infty$ limit, establishing that the simplified $\beta \rightarrow \infty$ expression from equation~(\ref{eq:P_inf}) provides an accurate representation of the full linear response theory in the high magnetic field limit.

With regard to the high magnetic field limit, recall that the expression for $k_\mx$ in equation~(\ref{eq:k_max}) is made without reference to the magnetic field. 
It is particularly uncertain if it applies when the gyroradius is smaller than the distance of closest approach, $r_c \lesssim r_L \Rightarrow \beta \gtrsim \Gamma^{-3/2}$~\cite{hu:02}. 
For $\Gamma = 10^{-3}$ the expression may be limited to the range $\beta \lesssim 3 \times 10^{4}$. 
Nevertheless, the figure establishes that the force smoothly transitions to the results of the $\beta \rightarrow \infty$ limit before, but near, the value at which the approximation for $k_\mx$ becomes uncertain. 
Indeed, the rate of convergence to this limit is governed by the parameter $\beta \Gamma^{3/2}$. 
Thus, one should be cautious that although the $\beta \rightarrow \infty$ limit expression corresponds to the result of linear response theory in this asymptotic limit, linear response theory itself may not be accurate for $\beta \gtrsim \Gamma^{-3/2}$ because of the limitation that the theory does not describe close collisions. 

The remainder of this paper explores consequences of the friction force computed in the strongly magnetized ($\beta \rightarrow \infty$) limit. 
One reason to focus on this limit is that it illustrates the basic physics processes associated with the influence of magnetization, with emphasis on the transverse force, in an extremum limit. 
It is also much less computationally expensive to solve equation~(\ref{eq:P_inf}) than to solve the full magnetized dielectric response function. 
This enables us to explore the phase dependence, wake potentials and trajectory calculations, which would be computationally difficult to do for an arbitrary magnetic field strength. 
Furthermore, even though the $\beta \rightarrow \infty$ limit is not quantitatively accurate for $\beta \ll \Gamma^{-3/2}$, it captures qualitative changes in both the stopping power and transverse force that are observed at lower values of $\beta$. 
Thus, it provides a good framework to explore the basic physical processes associated with strong magnetization. 

One example is shown in figure~\ref{fg:force_limits}. 
Panel (a) shows the stopping power ($-F_v$) when the test charge is either aligned parallel ($\theta = 0^\circ$) or perpendicular ($\theta = 90^\circ$) to the magnetic field. 
The transverse force vanishes in each of these limits. 
This panel demonstrates that in the high speed limit, $F_v$ is substantially diminished in comparison with the unmagnetized case when the test charge velocity is aligned with the magnetic field. 
This implies that a fast projectile aligned along the magnetic field in a strongly magnetized plasma has a much farther range than in an unmagnetized plasma. 
Magnetization effectively enhances the screening of the charge so that it collides less frequently with the background plasma. 
The same general character is observed when the charge moves perpendicular to the magnetic field, but to a much smaller degree. 
In contrast, magnetization is found to increase the stopping power for slow projectiles. 
The Bragg peak in both the parallel and perpendicular orientations is observed to transition to a lower speed. 
It is larger than the unmagnetized case for parallel orientation, but smaller than the unmagnetized case for perpendicular orientation. 

Panels (b) and (c) illustrate the phase dependence of the stopping power and transverse force. 
The Bragg peak of the stopping power is found to be largest in the direction parallel to the magnetic field ($\theta = 0^\circ$ or $180^\circ$) and to decrease monotonically with phase angle, reaching a minimum at perpendicular orientation ($\theta = 90^\circ$ or $270^\circ$). 
The position of the Bragg peak is also observed to shift to a lower speed as the phase angle transitions from a parallel to perpendicular orientation. 
An implication of these results is that the energy loss depends significantly on the orientation of the velocity vector with respect to the magnetic field. 

Panel (c) shows that the transverse component of the force has a periodic dependence on the phase angle. 
For speeds above a critical speed that is near the sonic speed ($v_c \simeq v_T$), it has a positive sign in the intervals $0^\circ-90^\circ$ and $180^\circ-270^\circ$, but a negative sign for $90^\circ-180^\circ$ and $270^\circ-360^\circ$. 
For speeds below the critical speed, the sign is opposite in each interval. 
As pointed out in section~\ref{sec:lrt}, the transverse force vanishes for both parallel and oblique orientations ($\theta = 0^\circ, 90^\circ, 180^\circ$ and $270^\circ$). 
A consequence is that the local maxima and minima are largest at oblique angles. 
These extrema values occur not at the $45^\circ$ angle shown in figures~\ref{fg:force_beta1} and \ref{fg:force_beta}, but rather at an angle of approximately $\pm 20^\circ$ from the direction parallel (or antiparallel) with the magnetic field. 
Comparing the transverse and stopping power components, the stopping power is the largest component of the force for speeds near the thermal speed (Mach 1). 
However, the transverse force decreases much less rapidly with speed than does the stopping power, and it becomes the dominant component of the force at high speed.

The phase dependence of the transverse force causes the gyroradius of fast particles ($v>v_c$) to increase and slow particles ($v<v_c$) to decrease. 
To better visualize this, figure~\ref{fg:sketch} shows a sketch of the force vectors in the $\vc{v}-\times$ coordinate representation when the angle lies in both the first ($0^\circ < \theta < 90^\circ$) and second ($90^\circ < \theta < 180^\circ$) quadrants. 
Consider a fast test charge. 
Panel (a) shows that a positive value for $F_\times$ in the first quadrant implies that the transverse force acts to reduce the component of the velocity parallel to the magnetic field ($v_\parallel$). 
Since the transverse force is perpendicular to $\vc{v}$, it does not influence total energy conservation. 
This implies that the loss of parallel energy due to the transverse force is compensated by an increase in the perpendicular energy, which increases the gyroradius. 
For comparison, panel (b) shows that a negative value for $F_\times$ is required to reduce the parallel velocity in the second quadrant. 
Since there is, indeed, a sign change of $F_\times$ from the first to second quadrant, this implies that the transverse force also acts to slow the parallel motion while increasing the gyroradius in this quadrant as well. 
Analogous arguments can be applied to the third and forth quadrants to show that the same conclusion is reached regardless of the phase angle. 
Section~\ref{sec:trajectories} further details the consequences of the transverse force on single particle motion.

\section{Wake potential\label{sec:wake}} 

\begin{figure}
\begin{center}
\includegraphics[width=8cm]{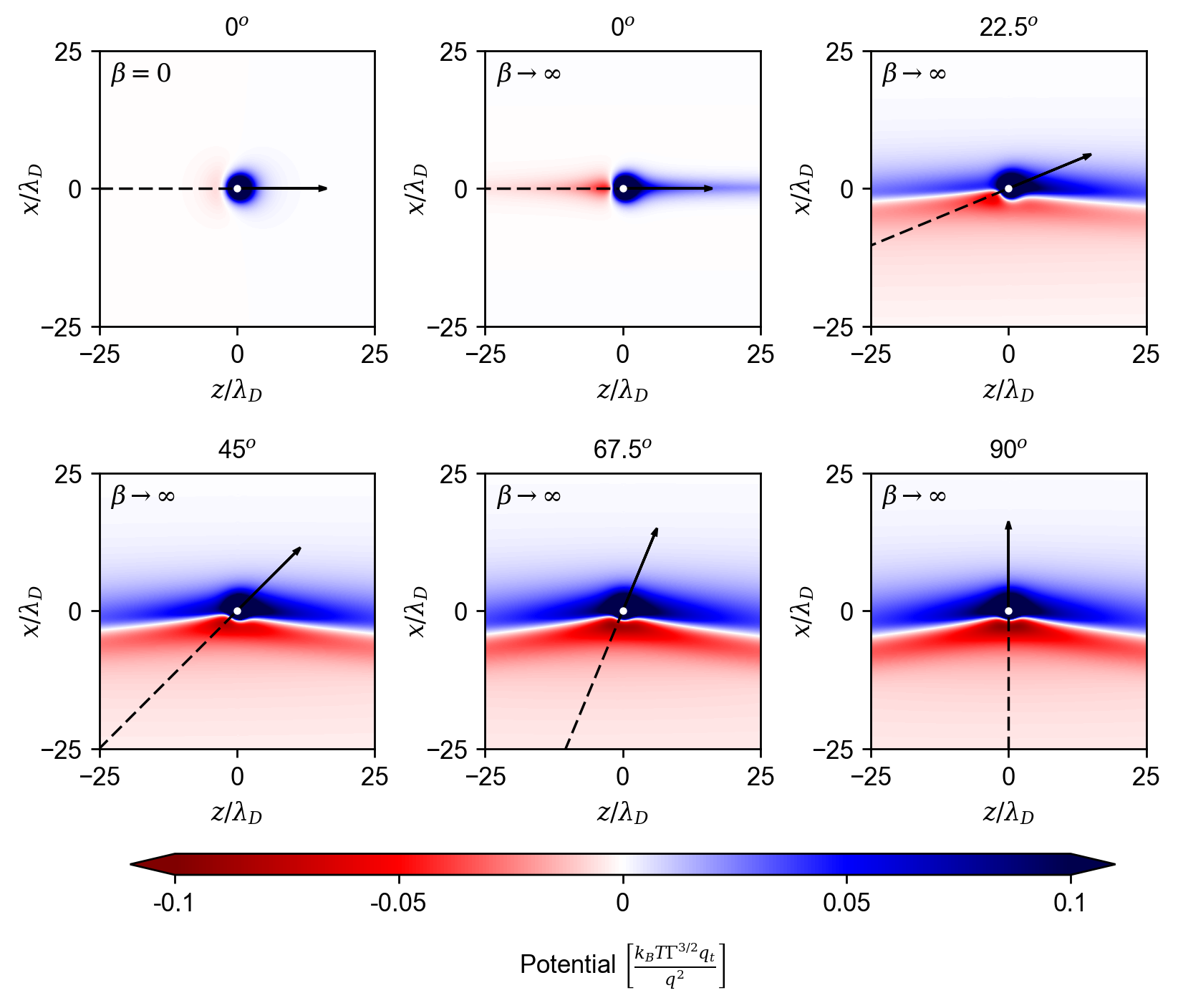}
\caption{Wake potential surrounding a test charge moving at a subsonic velocity ($M = 0.2$) at the indicated angle with respect to the magnetic field. The magnetic field is zero in the top left panel, while it is computed from the $\beta \rightarrow \infty$ limit from equation~(\ref{eq:beta_inf}) in the others.}
\label{fg:wake_angle_slow}
\end{center}
\end{figure}

\begin{figure}
\begin{center}
\includegraphics[width=8cm]{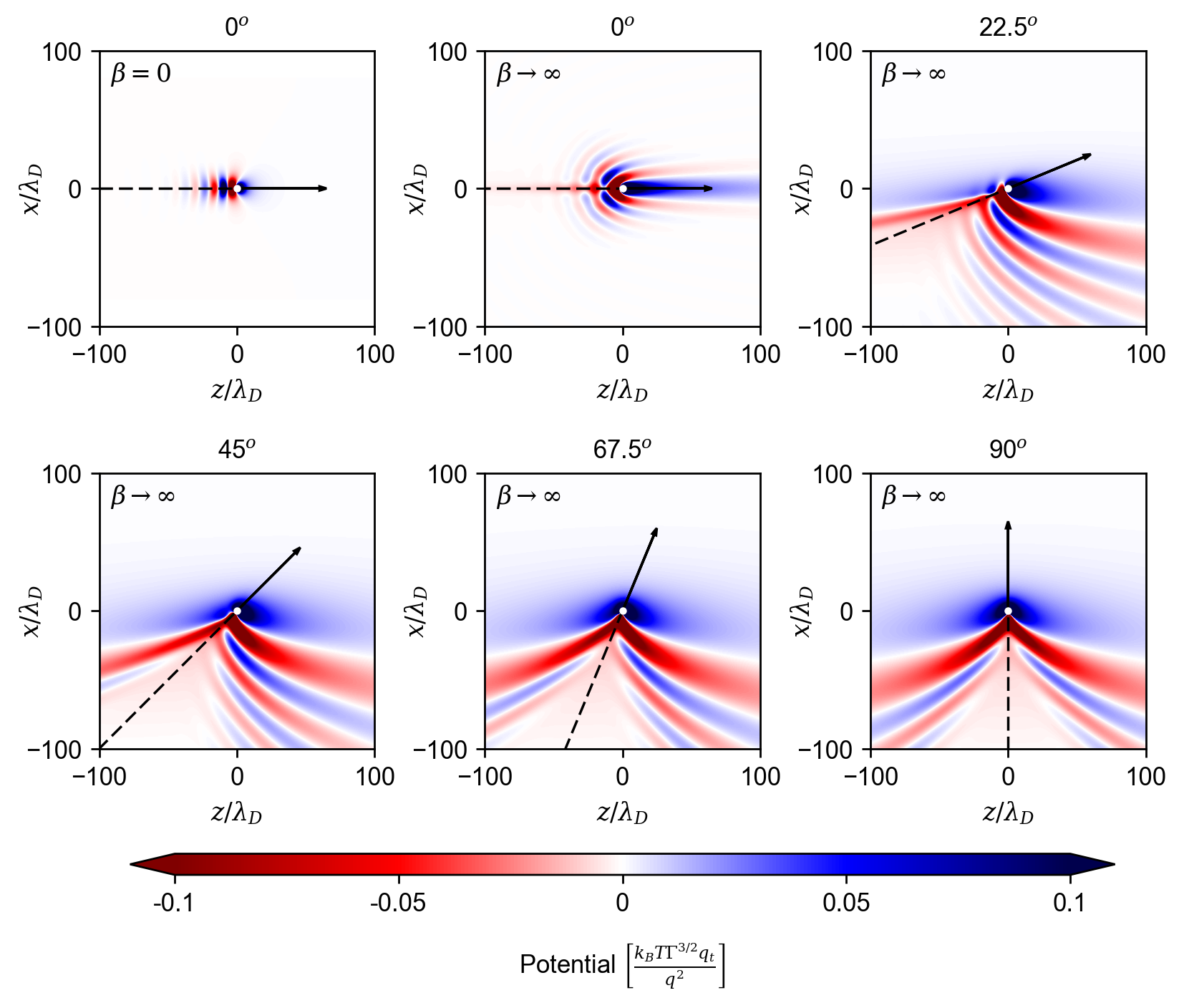}
\caption{Analogous to figure~\ref{fg:wake_angle_slow}, but for a supersonic test charge ($M = 2$).}
\label{fg:wake_angle_fast}
\end{center}
\end{figure}

The force on the test charge results from the induced electrostatic potential distribution of its wake. 
If the test charge is stationary ($v=0$), the wake potential is of the Debye-H\"{u}ckel form regardless of the strength of the magnetic field. 
Since this is a radially symmetric potential distribution, the associated force at the location of the test charge is zero. 
This is why both components of the force vector vanish as $v \rightarrow 0$. 
For a finite speed, deviations from radial symmetry arise that lead to a net force on the test charge. 
In an unmagnetized plasma, asymmetries arise along the direction of the velocity in such a way that the center of charge of the wake distribution lies behind the projectile; generating a drag force that slows the particle. 
Since the induced charge distribution is symmetric about the axis formed by the velocity vector, there is no transverse component of the force, as described in section~\ref{sec:lrt}. 
Figures~\ref{fg:wake_angle_slow} and \ref{fg:wake_angle_fast} show an illustration of the wake in an unmagnetized plasma for subsonic ($M = 0.2$) and supersonic ($M = 2$) speeds. 
These figures show the commonly expected form, including asymmetry in the direction of $\vc{v}$ and symmetry about its axis. 

\begin{figure*}
\begin{center}
\includegraphics[width=13cm]{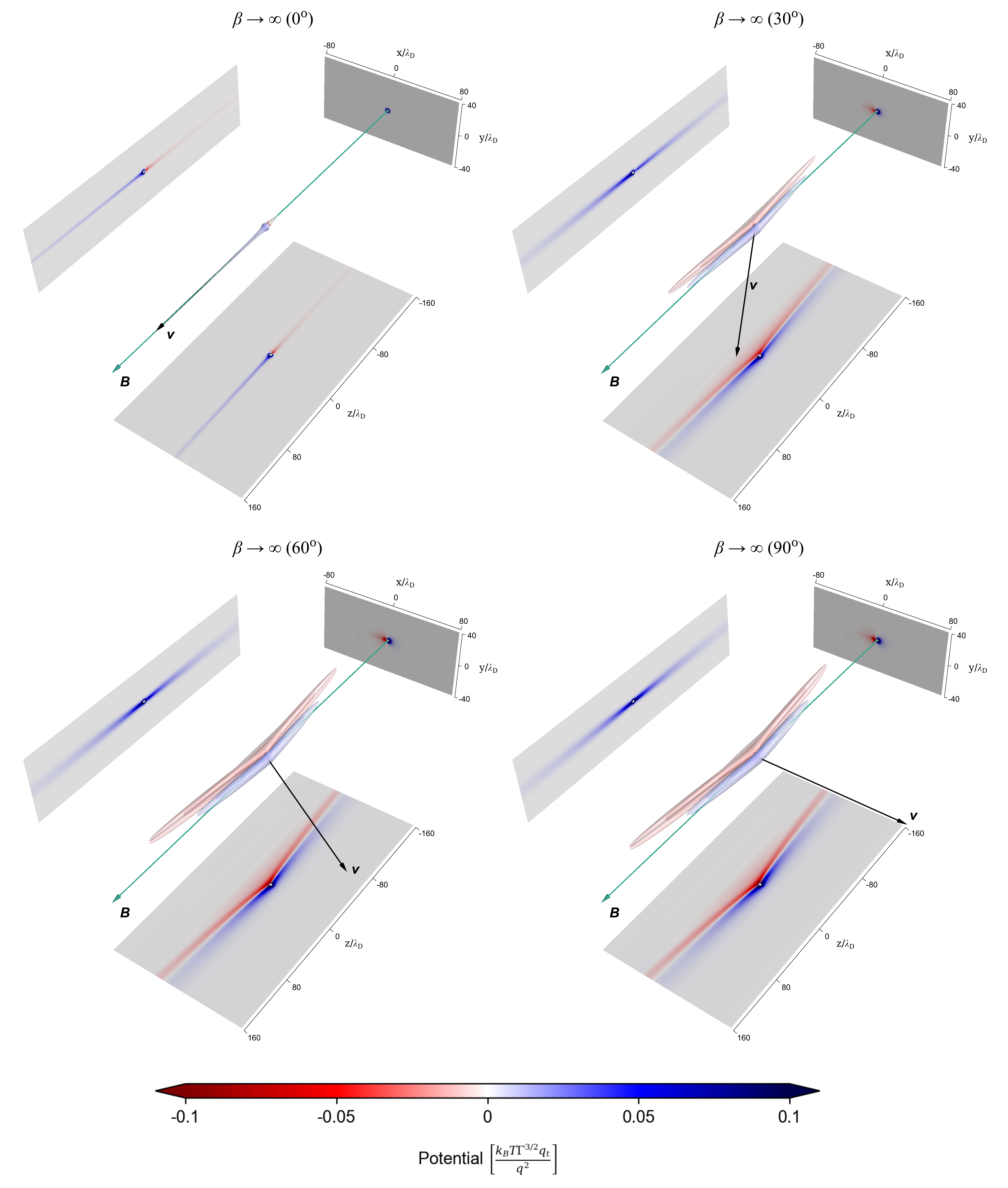}
\caption{Isosurfaces of the three-dimensional wake potential surrounding a test charge moving at a subsonic velocity ($M = 0.2)$ at the indicated angle with respect to the magnetic field. The potential was computed from the $\beta \rightarrow \infty$ limit from equation~(\ref{eq:beta_inf}) in all panels. Arrows represent the magnetic field vector (green) and velocity vector (black). }
\label{fg:3D_wakes_1}
\end{center}
\end{figure*}

This symmetry is broken in a strongly magnetized plasma when the projectile moves at an oblique angle with respect to the magnetic field. 
Figure~\ref{fg:wake_angle_slow} shows the wake potential for a subsonic speed ($M = 0.2$) as the angle rotates from $0^\circ$ to $90^\circ$. 
At parallel incidence ($\theta = 0^\circ$) the magnetic field causes a steepening of the potential profile behind the charge and an extension of the wake to much further distances both ahead and behind the particle. 
These changes lead to the increase in stopping power observed at subsonic speeds when comparing the blue and black lines in figure~\ref{fg:force_limits}. 
As the angle rotates, the wake is observed to broaden over a much larger region of space but primarily along the magnetic field direction. 
This alignment of features of the wake with the magnetic field causes it to form an asymmetry with respect to the velocity vector. 
It is this asymmetry that leads to the transverse force observed in figures~\ref{fg:force_beta1}--\ref{fg:force_limits}. 
Although the wake is stretched into the magnetic field direction, it is not symmetric about the magnetic field axis at oblique incidence. 
Symmetry returns to the velocity axis when it is perpendicular to the magnetic field ($90^\circ$). 
This is why the transverse force vanishes at perpendicular incidence, as shown in figure~\ref{fg:force_limits}, which is a simple analytic limit of equation~(\ref{eq:f_times}) that was discussed in section~\ref{sec:lrt}.

Figure~\ref{fg:wake_angle_fast} shows a similar set of wake potentials, but for a supersonic test charge ($M = 2$). 
Qualitative features of this wake potential are similar to figure~\ref{fg:wake_angle_slow}. 
Again, symmetry is observed about the velocity vector for parallel ($\theta = 0^\circ$) and perpendicular ($\theta = 90^\circ$) orientations, but not at oblique angles. 
The magnetic field causes the wake to spread over a much broader region of space. 
The fast particle wake has a much more detailed fine structure, including more oscillations, than for a slow particle. 
We also note that the fore-wake extends very far in front of the charge when it is aligned parallel to the magnetic field, but it becomes much less significant at oblique, and even perpendicular, incidence. 
Non-trivial curving of the wake behind the particle is also observed. 
This is associated with the way in which the Lorentz force acts on the linear plasma dielectric. 

\begin{figure*}
\begin{center}
\includegraphics[width=13cm]{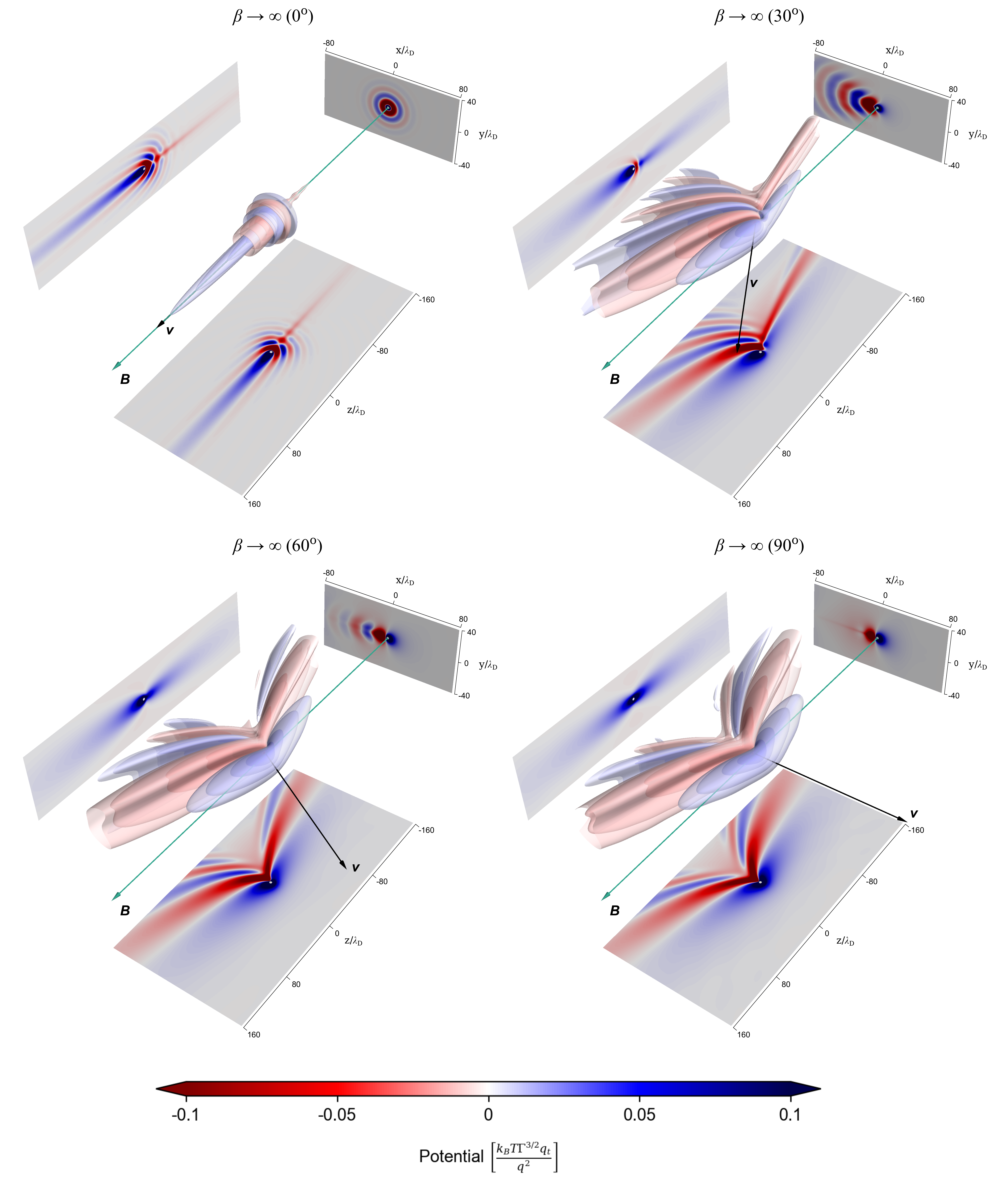}
\caption{Analogous to figure~\ref{fg:3D_wakes_1}, but for a supersonic test charge ($M = 2$). }
\label{fg:3D_wakes_2}
\end{center}
\end{figure*}

The wake potential can also be visualized as isosurfaces in 3D. 
Figure~\ref{fg:3D_wakes_1} shows such a 3D representation for a slow particle, corresponding to the same subsonic ($M = 0.2$) speed as figure~\ref{fg:wake_angle_slow} for a range of angles between $\theta = 0^\circ$ and $90^\circ$. 
Although the steering caused by the asymmetry of the wake can be seen in 2D, the 3D representation shows another dimension of how the magnetic field influences the wake. 
Most notably, the wake is observed to be much more compressed in the plane normal to $\vc{v}$ and $\vc{B}$ (the $\hat{\vc{n}}$ direction) than it is in the $\vc{v}-\vc{B}$ plane. 
In this dimension (the $y$ direction in this figure), the magnetic field appears to have a much less significant influence, such that the screening length still appears to be characterized primarily by the Debye length (as it is when $\beta =0$ or $\theta = 0^\circ$). 
It is interesting to note that the projection of the wake potential onto the $y-z$ plane is entirely of one sign at sufficiently oblique angles, whereas the projection onto the $x-y$ plane shows a sign change between the potential in the fore-wake and behind the test charge. 
The figure also displays the expected symmetry about the $x-z$ plane. 
This symmetry illustrates why the force on the particle lies in the $\vc{v}-\vc{B}$ plane, i.e., there is no $F_n$ component to the force, as expected from section~\ref{sec:lrt}.  

Figure~\ref{fg:3D_wakes_2} shows an analogous 3D representation of the wake for a supersonic test charge ($M = 2$). 
Many features of the wake are similar to that of the slow particle, including the comparatively shorter screening length in the direction perpendicular to the $\vc{v}-\vc{B}$ plane ($y$ direction) than in the plane. 
The expected symmetry about this plane is also clear.  
Several distinguishing qualities are also observed. 
The screening length is generally larger for faster particles, so the isosurfaces extend much further in space. 
Much more detailed structure is also observed, as in 2D. 
Again, the projection onto the $y-z$ plane is observed to be of one sign for sufficiently large angles ($\theta = 60^\circ$ and $90^\circ$), but to have fine-scale features of each sign at smaller angles. 
The projection onto the $x-y$ plane makes significant qualitative changes as $\theta$ varies. 
For parallel incidence ($\theta = 0^\circ$) it consists of concentric circles, whereas a trailing conical wake with semicircular oscillations is observed for the gradual oblique angle ($\theta = 30^\circ$). 
As the angle increases further ($\theta = 60^\circ$), the angle of the conical envelope continues to shallow as does the depth of the oscillating potential trailing the particle. 
Finally, at perpendicular incidence the wake is a bipolar structure that is qualitatively similar to what is observed for a slow particle, but with a still extended range. 
 
These figures illustrate that the Lorentz force influences the wake potential, and by extension, the induced force on the particle. 
Next, we consider how that induced force ultimately influences the trajectory of a test charge as it traverses the magnetized linear dielectric medium. 

\section{Trajectory\label{sec:trajectories}} 

This section considers an example single particle trajectory in a strongly magnetized plasma computed from equation~(\ref{eq:eom}). 
Because of its relevance to fusion, we consider a He$^{2+}$ ion slowing on electrons. 
To emphasize the influence of the transverse force, the trajectory is computed both from the complete description and by arbitrarily setting the transverse force to zero ($F_\times = 0$). 
Both components of the friction force are computed using the $\beta \rightarrow \infty$ description from equation~(\ref{eq:P_inf}). 

\begin{figure}
\begin{center}
\includegraphics[width=8cm]{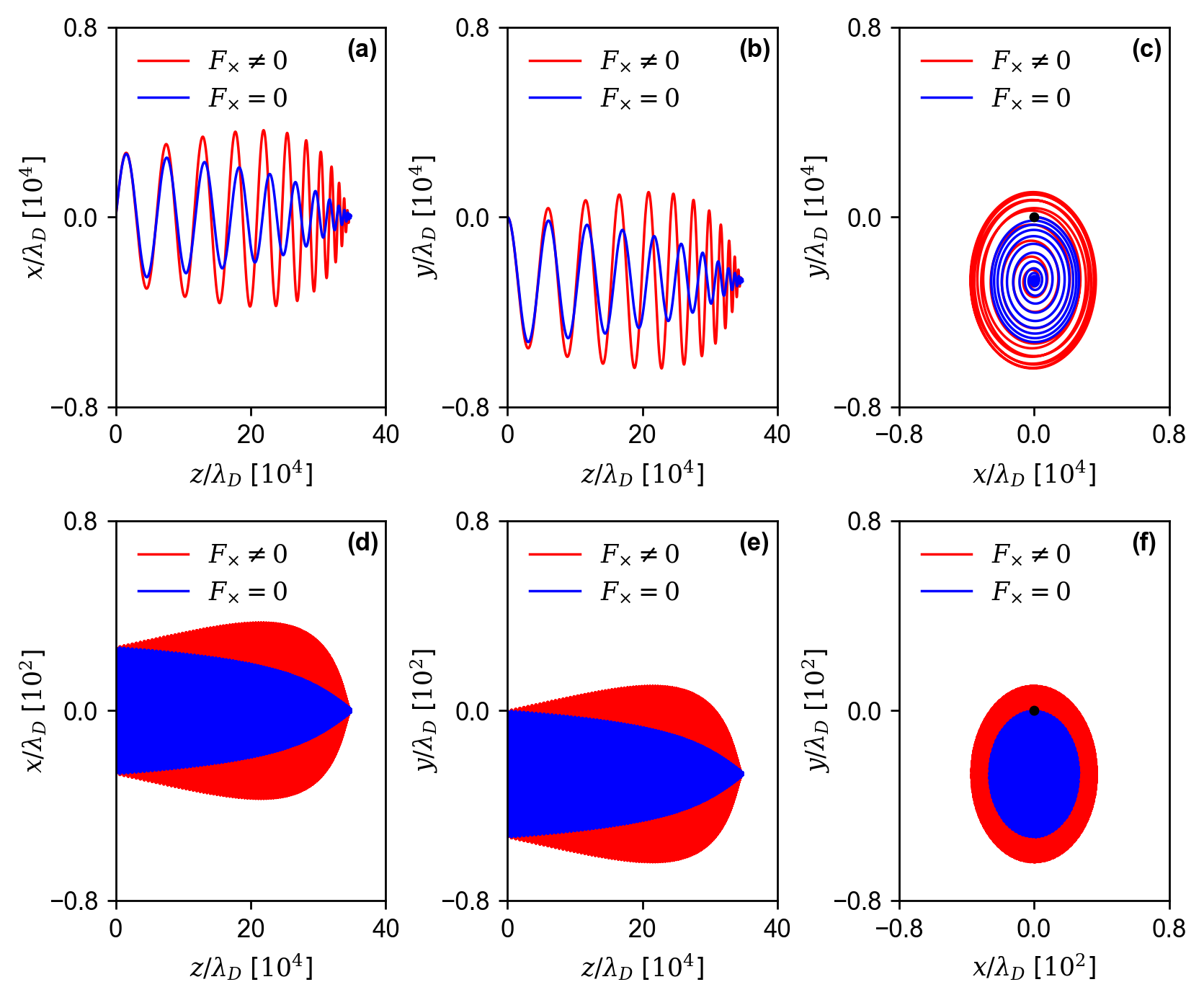}
\caption{Trajectory of a supersonic ($M = 2$) He$^{2+}$ ion initially oriented at a $15^\circ$ angle with respect to the magnetic field in a magnetized electron plasma with $\Gamma = 0.1$. In the top row (a)-(c) $\beta = 1$, while in the bottom row (d)-(f) $\beta = 100$. In both cases, the friction force components $F_v$ and $F_\times$ were computed from the $\beta \rightarrow \infty$ limit of the dielectric function from equation~(\ref{eq:beta_inf}). The red lines show the trajectory computed using the full friction force, including the transverse component $F_\times$, whereas the blue lines were obtained by artificially setting $F_\times = 0$. The gyrofrequency is so large in panels (d)-(f) that the colored regions appear solid. The black dot in panels (c) and (f) indicate the initial ion position.}
\label{fg:trajectory_1}
\end{center}
\end{figure}

Figure~\ref{fg:trajectory_1} shows the trajectory of an initially supersonic particle ($M = 2$) at an angle of $\theta = 15^\circ$ with respect to the magnetic field in a background plasma with coupling strength $\Gamma = 0.1$. Here the particle begins at the origin, the magnetic field vector has been aligned along the $z$-axis, and the initial particle velocity vector lies in the $x$-$z$ plane: i.e. the initial particle velocity components are $v_x = v\sin\theta$, $v_z = v\cos\theta$, and $v_y = 0$.
The top row shows a solution using a value of $\beta = 1$ to compute the Lorentz force term in equation~(\ref{eq:eom}). 
It should be noted that using $\beta = 1$ for the Lorentz force is not a large enough value to be consistent with the $\beta \rightarrow \infty$ limit used to compute the friction force. 
This solution is provided simply to illustrate qualitatively how the friction force terms influence the trajectory over a gyrorbit, which is very difficult to see at a higher $\beta$ value because the gyromotion is so rapid. 
The bottom row in figure~\ref{fg:trajectory_1} shows the solution using a value of $\beta = 100$. 
This value of $\beta$ is consistent with the $\beta \rightarrow \infty$ solution for the friction force at this value of~$\Gamma$.

Textbook solutions of single particle motion include only the Lorentz force term and the stopping power $F_v$ (blue lines in figure~\ref{fg:trajectory_1}). 
Because the stopping power acts only along the direction of the velocity, it causes the gyoradius to monotonically decrease as the particle slows. 
It eventually stops at a range of approximately $3.5 \times 10^5 \lambda_D$ in this example. 
The rate at which the gyroradius decreases is determined by the speed of the particle at that instant, according to the force illustrated in figure~\ref{fg:force_limits}. 
Similar behavior is observed for both $\beta$ values, with the only observable difference being the gyrofrequency. 
The envelope functions describing the gyroaveraged motion appear indistinguishable. 

In contrast, when the transverse force is included in the computation the gyroradius is observed to \emph{increase} for the majority of the evolution, before decreasing rapidly near the end of its range. 
This behavior is a result of the combined influence of the transverse force and the stopping power. 
Because of the phase dependence shown in figure~\ref{fg:force_limits}, the transverse force component in equation~(\ref{eq:eom}) always acts to decrease the speed of the test charge parallel to the magnetic field ($v_\parallel$) when the speed is above the critical speed ($v>v_c$). 
This causes the parallel velocity to decrease more rapidly when the transverse force is included than when it is not. 
This effect can be clearly seen in the top row of figure~\ref{fg:trajectory_1}, where the phase of the solution including the transverse force is observed to lag that of the solution excluding the transverse force. 
Since the transverse force is perpendicular to the velocity of the test charge, it does not contribute to energy loss (slowing). 
Energy conservation implies that the energy lost from the parallel direction is redirected to an increase in the perpendicular speed ($v_\perp$). 
The increase in the perpendicular speed leads to the observed increase of the gyroradius $r_{c,t} = v_\perp/\omega_{c,t}$. 
These expectations can also be seen by splitting equation~(\ref{eq:eom}) into components parallel and perpendicular to the magnetic field.  

Stopping power acts to slow the test charge simultaneously with this action of the transverse force. 
This reduces both the parallel and perpendicular speeds as the particle evolves, which nominally acts to decrease the gyroradius. 
The net increase in the gyroradius for early times occurs because the transverse force exceeds the stopping power for sufficiently fast projectile speeds. 
Eventually, the stopping power causes the projectile speed to decay sufficiently that the sign of the transverse force switches; see figure~\ref{fg:force_limits}. 
After this occurs, both the transverse force and stopping power components act to reduce the gyroradius. 
This results in a much steeper decrease of the gyroradius toward the end of the evolution when the transverse force is included than when it is not. 

Because the transverse force acts to slow the parallel velocity, it shortens the range of the projectile. 
This is a very small change at the conditions plotted in figure~\ref{fg:trajectory_1}. 
However, the difference between the curves with and without the transverse component is larger when the initial parallel velocity of the projectile is larger. 
In fact, the transverse force has a significant influence on the range of the test charge when the initial parallel speed is much larger than the perpendicular speed. 
Aside from the fundamental physics interest, these results may be relevant to many applications in which the slowing of a charge in a strongly magnetized plasma is important, including energy deposition of fusion products~\cite{sigm:71} and runaway electron physics in fusion~\cite{paz:14}, as well as diagnostics in high energy density plasma experiments~\cite{li:06}.

\section{Conclusions} 

Linear response theory predicts that strong magnetization not only influences stopping power, but also introduces a transverse component to the friction force on a test charge moving through a plasma. 
The physical process responsible for this behavior is the influence of the Lorentz force on the wake induced by the test charge. 
Magnetization qualitatively changes the wake. 
It causes a reduced charge density directly behind the charge as well as finely structured oscillations for any orientation between the velocity and magnetic field vectors. 
When the velocity is at an oblique angle with respect to the magnetic field, the wake is asymmetric with respect to the velocity vector.  
The induced electric field associated with the charge density in this wake generates a friction force, and the asymmetric component leads to the transverse component of this force. 

The transverse force is observed to change sign depending on the speed of the test charge as well as the angle between the velocity and magnetic field vectors. 
The phase dependence implies that the transverse force acts to decrease the parallel speed, and increase the perpendicular speed, of a fast (approximately supersonic) test charge; causing the gyroradius to increase. 
Due to the sign change at low speeds, it conversely acts to increase the parallel speed, and decrease the perpendicular speed of a slow (approximately subsonic) test charge; causing the gyroradius to decrease. 
Because the transverse force is perpendicular to the velocity vector, it does not directly contribute to the energy loss rate of the test charge. 
However, the stopping power acts simultaneously to slow the particle throughout the orbit. 
This causes the particle to eventually stop, and the transverse force contribution to change in time correspondingly. 
For an initially fast test charge, the combined influence of the transverse force and stopping power is that the gyroradius increases until the speed decreases below approximately the thermal speed, at which point the gyroradius decreases rapidly as the particle slows to a stop.  

These fundamental physical effects on single particle motion may be of interest to several applications such as slowing of fusion products and runaway electron generation, as well as to understanding other transport processes in strongly magnetized plasmas. 
Considering a test charge slowing on a OCP, these effects are observed to onset when $\beta \gtrsim 1$, and to asymptote to a simplified strong magnetic field limit as $\beta$ approaches a value of $\Gamma^{-3/2}$. 
Many plasma experiments reach regimes at which $\beta$ can be greater than unity. 
For instance, electrons in many magnetic fusion energy experiments can lie in this regime. 
This suggests that strong magnetization effects could be relevant to some processes in these experiments. 
However, it should be emphasized that the asymptotic regime where $\beta$ approaches $\Gamma^{-3/2}$ is much less common in hot dilute (weakly coupled $\Gamma \ll 1$) plasmas, such as magnetic fusion energy experiments.  
Instead, this limit is most relevant to moderately coupled plasmas ($\Gamma$ only slightly less than unity), such as nonneutral plasmas~\cite{glin:92}, ultracold neutral plasmas~\cite{zhan:08}, and dense plasmas~\cite{slut:10}. 
It is also relevant to systems where the test charge is very large, such as magnetized dusty plasma experiments~\cite{thom:12}. 
The linear response theory presented here treated weakly coupled plasmas, and would have to be modified to address moderate or strong Coulomb coupling. 
Nevertheless, the basic physics mechanism of the transverse force is expected to also apply in these situations. 

\section*{Acknowledgments}

This material is based upon work supported by the U.S. Department of Energy, Office of Science, Office of Fusion Energy Sciences under Award Number DE-SC0016159 and Air Force Office of Scientific Research under Award No.~FA9550-16-1-0221.


\end{document}